\documentstyle[prb,aps,graphicx,multicol,amssymb]{revtex}

\draft

\newcommand{\bN}{{\bf 0}}
\newcommand{\br}{{\bf r}}

\newcommand{\deltan}{{\delta n}} 
\newcommand{\grad}{{\mbox{\boldmath $\nabla$}}}
\newcommand{\tDelta}{{\widetilde \Delta}}

\begin{document}

\title{Pinning of stripes in cuprate superconductors}

\author{Simon Bogner and Stefan Scheidl} 

\address{Institut f\"ur Theoretische Physik, Universit\"at zu K\"oln,
  Z\"ulpicherstr. 77, D-50937 K\"oln, Germany}

\date{\today} 

\maketitle

\begin{abstract} 
  We examine the effects of disorder on striped phases in
  high-temperature superconductors and related materials.  In the
  presence of quenched disorder, pinning by the atomic lattice --
  which might give rise to commensuration effects -- is irrelevant for
  the stripe array on large length scales.  As a consequence, the
  stripes have divergent displacement fluctuations and topological
  defects are present at all temperatures.  Therefore the positional
  order of the stripe array is short ranged, with a finite correlation
  length even at zero temperature.  Thus lock-in phenomena can exist
  only as crossovers but not as transitions.  In addition, this
  implies the glassy nature of stripes observed in recent experiments.
\end{abstract}


\begin{multicols}{2}

\section{Introduction}

During the last decade evidence emerged from theoretical
work\cite{Schulz89,Zaanen+89,Emery+90} and from experiments on
cuprates\cite{Cheong+91,Mason+92,Tranquada+95,Dai+98} and closely
related nickelates\cite{Hayden+92,Chen+93,Tranquada+94,Lee+97} for the
existence of striped structures within the $M$O$_2$ ($M=$ Cu, Ni)
planes.  These stripes are highly correlated states of holes which are
introduced into the planes by doping and which order into a
unidirectional\cite{Mook+00} charge-density wave (CDW, wave length
$a$) that may be accompanied by a simultaneous spin-density wave of
period $2a$ in the sublattice magnetization of the antiferromagnetic
metallic spins.\cite{Tranquada+94,Kivelson+96,Zachar+98}
Qualitatively, one may think of stripes as parallel strings of holes
that constitute an antiphase boundary for spin order.

Particular interest in these stripes arises from the possible
interplay\cite{Emery+93} between these stripes and superconductivity.
Thereby it is important to distinguish between ``dynamic'' and
``static'' stripes.  While there is evidence that superconductivity
can coexist with both dynamic\cite{Cheong+91,Mason+92,Mook+99} and
static\cite{Tranquada+97,Kimura+99} stripes, static stripes tend to
suppress superconductivity,\cite{Crawford+91,Nakamura+92,Emery+97} in
contrast to dynamic stripes.  Therefore the study of the structure and
dynamics of stripes is of principal importance.

Various phenomenological pictures have been developed for the
theoretical description of stripes.  While charge and spin order are
naturally described within a Landau theory,\cite{Zachar+98} the aspect
that stripes act as magnetic domain walls suggests to describe them as
string-like objects.\cite{Zaanen+96,MSmith+98} In the ideally ordered
case these strings form a periodic array. Dynamic fluctuations are
generated by thermal and quantum effects, whereas potentials tend to
suppress dynamic fluctuations while they may reduce or increase static
conformations of the stripes.

In such phenomenological models, the crystalline structure of the
underlying atomic lattice (period $b$) has to be taken into account by
a periodic potential, which tends to increase the positional order of
the stripe array since is can be the source of lock-in
effects.\cite{Tranquada+95,Nayak+96,Emery+97} On the other hand, the
spatially inhomogeneous distribution of dopants provides a disorder
potential for the stripes because of the (screened) Coulomb
interaction between dopants and holes. For low enough temperatures,
these dopants can be considered as quenched.

Pinning by the periodic potential is of particular interest since
lock-in effects might explain the special role of certain values for
the stripe spacing.  In the cuprate system La$_{2-x}$Sr$_x$CuO$_4$,
the mismatch $\delta$ of magnetic Bragg satellite peaks has a nonzero
value only beyond a threshold doping $x_c \approx 0.04$, where $\delta
\approx x$ holds. Eventually a saturation value $\delta=\frac 18$ is
observed for $x \gtrsim 0.12$.\cite{Yamada+98} Since the mismatch
$\delta$ is related to the lattice and CDW periods via $\delta=b/2a$,
it allows for a natural explanation\cite{Tranquada+95} of the
``$x=\frac 18$-problem'':\cite{Moodenbaugh+88} Since at $\delta=\frac
18$ the periods of the CDW and the Cu spacings have an integer ratio
$p:= a/b=4$ this saturation could be a commensuration
effect.\cite{note_comm} Similarly, the nickelate system
La$_{2-x}$Sr$_x$NiO$_{4+y}$ shows anomalous thermodynamic behavior at
the values $p=2, 3$\cite{Cheong+94,Ramirez+96} which seem to be stable
over certain respective ranges of the hole concentration
$x+2y$.\cite{Chen+93} Even more, evidence was
reported\cite{Tranquada+94} for plateaus of the mismatch as a function
temperature at rational values $\delta=\frac3{22}$, $\frac5{36}$ in
La$_2$NiO$_{4.125}$.

On the other hand, the stripe array can also be pinned by
disorder.\cite{Tranquada+99,Du+00} Therefore it is important to take a
closer look at the competitive pinning by the periodic atomic lattice
and by disorder in order to understand to what extent lock-in effects
can persist.  A first step in this direction was made by Hasselmann
{\em et al.}\cite{Hasselmann+99} who focus on a {\em single} stripe.
However, since a single stripe and a stripe array differ in
dimensionality one expects qualitatively distinct behavior of the
response of the system to disorder.

The purpose of this paper is to determine the effects of periodic and
disorder potentials on the structural order of the two-dimensional
stripe array on large scales.  Quantum fluctuations turn out to be
irrelevant in the presence of disorder, i.e. the system behaves
essentially as a classical stripe array.  We find that a locked state
with long-ranged positional order can exist only in the absence of
disorder. Disorder always leads to unlocking and a state with
short-ranged positional order where dislocations proliferate.

The outline of this article is as follows. In Sec. \ref{sec_model} we
establish an elastic model for the quantum stripe array with periodic
and pinning potentials.  The effects of these potentials are discussed
in Sec.  \ref{sec_class} in the classical limit.  In Sec.
\ref{sec_quant} we demonstrate the irrelevance of quantum
fluctuations.  Our conclusions are drawn in Sec. \ref{sec_disc} where
we discuss the relation of our work to previous work, the role of
topological defects and implications for experiments.

\section{The Model}
\label{sec_model}

We describe the stripe system as an array of interacting quantum
strings.  The strings are assumed to be aligned in $y$ direction and
to have an average spacing $a$ in $x$ direction.  In the following we
ignore topological defects in the array, the role of which will be
discussed in Sec. \ref{sec_disc}.  Then the stripe array can be
considered as an elastic system. The displacement field $u$ represents
a bosonic collective mode of the electron system.  Its fluctuations
are governed by a (``reduced'') dynamic action
\cite{Zaanen+96,MSmith+98,Hasselmann+99}
\begin{eqnarray}
  {\cal S} &=& \frac 1 \hbar 
  \int_0^{\hbar/T} d\tau  \left\{ \int d^2r \ 
  \frac \mu 2 (\partial_\tau u)^2  + {\cal H} \right\}.
\label{def.S}
\end{eqnarray}
The action is formulated using the imaginary time (we set $k_{\rm
  B}=1$).  We identify $\br=(x,y)$, $\mu$ is a mass density and the
Hamiltonian ${\cal H}$ has a contribution from the elastic energy
\begin{eqnarray}
  {\cal H}_{\rm el} &=& \int d^2r \ \frac \gamma 2 (\grad u)^2
\end{eqnarray}
with a stiffness constant $\gamma$ which includes the line tension of
the strings as well as their interaction.  Besides a contributions of
entropic nature,\cite{Zaanen00} the main contribution to this
interaction will stem from the Coulomb interaction between the
stripes.  A further contribution arises from a crystal field that
aligns the stripes in $y$ direction.  {\em A priori}, the stiffness
can be anisotropic with an elastic energy density $\propto \gamma_x
(\partial_x u)^2 + \gamma_y (\partial_y u)^2$.  To simplify the
analysis, such an anisotropy can be removed by rescaling the $y$
coordinate. Then the effective isotropic stiffness constant is related
to the original anisotropic constants through
$\gamma=\sqrt{\gamma_x\gamma_y}$.  Note that $\gamma_x$ is dependent
on the stripe spacing $a$, i.e., on doping.  With increasing distance
$a$ between the stripes $\gamma$ will shrink.

We will examine the coupling of the stripe array to the periodic
potential $U(x)$ generated by the atomic structure as well as to a
random potential $V(x,y)$ due to the interaction between the holes and
the dopants which may be considered quenched at low temperatures.  The
corresponding energy contributions read
\begin{eqnarray}
  {\cal H}_U = \int d^2r \ \rho(\br) U(x) , \quad 
  \label{H_U}
  {\cal H}_V = \int d^2r \ \rho(\br) V(x,y)
  \label{H_V}
  \label{H_pin}
\end{eqnarray}
in terms of the stripe density
\begin{eqnarray}
  \label{density}
  \rho(\br,\tau) \simeq \frac 1a \Big\{ \sum_m 
  e^{i Q_m [x-u(\br,\tau)] }  - \partial_x u(\br,\tau) 
  \Big\}
\end{eqnarray}
where $Q_m:={2\pi m} /a$ are reciprocal lattice vectors of the stripe
array.  The elastic and disorder pinning energies of the stripe array
are similar to those of vortex lines in planar type-II
superconductors.  A recent review of the latter system can be found in
Ref. \onlinecite{Nattermann+00}.

$U$ is assumed to be periodic, $U(x)=U(x+b)$, with a period $b<a$ (the
modulation along the stripes is negligible for our purposes).  For
simplicity we take $U(x)$ as an even function (this restriction is for
the simplicity of our analysis but not essential for the results)
\begin{eqnarray}
  U(x)=-\sum_{n\geq1} U_n \cos(p Q_n x)
\end{eqnarray}
with $p>1$.
We assume that the random potential $V(x,y)$ is Gaussian distributed
with zero average and a variance
\begin{eqnarray}
  \overline{V(\br) V({\bf 0})}&=&\frac{\Delta}{\sqrt{2 \pi} \ \xi}
  e^{-x^2/2\xi^2} \delta(y)
\end{eqnarray}
with a correlation length $\xi$ and a weight $\Delta$. 

Subsequently we will establish the global phase diagram for the
total system with a partition sum
\begin{eqnarray}
  \mathcal{Z}&=&\int{\mathcal{D}}[u] \ e^{-S}.
  \label{def.Z}
\end{eqnarray}
Since the system without pinning provides an important reference point
we start with a brief discussion of thermal and quantum fluctuations
of the displacement.  There is a characteristic length scale $\ell_T
:= \sqrt{\hbar^2 \gamma/T^2 \mu}$ beyond which thermal fluctuations
dominate over quantum fluctuations.  A related temperature scale
$T_a:= \sqrt{\hbar^2 \gamma/a^2 \mu}$ is defined by the coincidence
$\ell_T=a$. In terms of these scales, the displacement fluctuations in
systems with a large size $L \gg \ell_T$ are obtained as
\begin{eqnarray}
  \langle u^2 \rangle \simeq \left\{
    \begin{array}{ll}
      \displaystyle
      \frac T {2 \pi \gamma} \ln \frac L a  & \textrm{ for } T \gg T_a,
      \\
      \displaystyle
      \frac T {2 \pi \gamma} \ln \frac L {\ell_T} + 
      \frac{\hbar}{2 a \sqrt{\pi \gamma \mu}}  & \textrm{ for } T \ll
      T_a .
    \end{array}
  \right.
\label{corr_u_el}
\end{eqnarray}
Thus, while the unpinned stripe array is flat (i.e., $\langle u^2
\rangle$ is finite for $L \to \infty$) at $T=0$, it is logarithmically
rough (i.e., $\langle u^2 \rangle \propto \ln L$ for $L \to \infty$)
at any finite temperature.

\section{Classical limit ($\hbar=0$)}
\label{sec_class}

For the analysis of the effects of the potentials it is convenient to
examine the various limiting cases defined by the relative strength of
thermal fluctuations, quantum fluctuations, periodic pinning and
disorder pinning.  We start from the consideration of the classical
limit $\hbar \to 0$ acting on the $\hbar$ appearing explicitly in Eq.
(\ref{def.S}) but not on possible implicit dependences of other model
parameters.  In this limit temporal fluctuations become negligible and
one has to examine the system governed by the Hamiltonian
\begin{eqnarray}
\label{Hamiltonian}
  {\cal H} &=& {\cal H}_{\rm el} +  {\cal H}_U +  {\cal H}_V .
\end{eqnarray}
In the absence of the potentials $U$ and $V$ thermal fluctuations lead
to an average displacement that diverges logarithmically with the
system size $L$ [cf. Eq. (\ref{corr_u_el}) for $T_a=0$] which means
that the stripe structure has only quasi-long-range order in the
position of the stripes.

\subsection{Periodic potential only ($V=0$)}
\label{sec_U}

\begin{figure}
\includegraphics[width=0.9\linewidth]{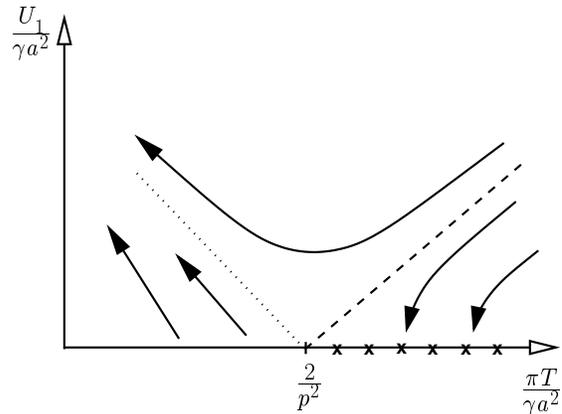}
\caption{Schematic representation of the renormalization-group flow
  near the roughening transition according to Eqs.
  (\ref{flow_roughening}). The line of crosses represents fixed points
  where the periodic potential is irrelevant and the stripes are not
  locked. The dashed line is the phase boundary between the locked and
  unlocked phase.  Arrows indicate the RG flow.}
\label{fig_roughening}
\end{figure}

To analyze the relevance of a periodic pinning potential we focus on
commensurabilities of low order with integer $p$.  In this case the
stripe structure can lock into the periodic potential at low
temperatures while it unlocks at large temperatures.  The transition
between these two states is analogous to the roughening transition of
crystal surfaces.  We follow the standard analysis of the roughening
transition (see Ref.  \CITE{Nozieres92} and references therein) in
order to obtain the transition temperature $T_{\rm R}$.  Combining
Eqs. (\ref{H_U}), (\ref{density}) and (\ref{corr_u_el}) we find an
average potential energy
\begin{mathletters}
\label{H_U.scal}
\begin{eqnarray}
  \langle {\cal H}_U \rangle &\simeq&
  - L^2 \frac 1 a \sum_{n \geq 1} U_n
  \exp \left( -p^2 Q_n^2 \langle u^2 \rangle /2 \right)
  \label{H_U.av}
  \\
  &\simeq&
  - L^2 \frac 1 a \sum_{n \geq 1} U_n
  \left( \frac L a \right)^{-p^2 Q_n^2 T/4 \pi \gamma}
\end{eqnarray}
\end{mathletters}
for an infinitesimally weak periodic potential.  The lowest harmonic
$n=1$ gives the most relevant contribution to this energy.  The
stripes are locked when the average potential energy does not vanish
in the limit $L \to \infty$, i.e., for temperatures below
\begin{eqnarray}
  T_{\rm R}= \frac {2 \gamma a^2}{\pi p^2}.
\label{def.T_R}
\end{eqnarray}
The transition temperature increases with increasing strength of the
potential since $\gamma$ is renormalized to larger values.  The
effective parameters on large length scales $L=ae^l$ are described by
renormalization group (RG) flow equations\cite{Nozieres92}
\begin{mathletters}
\label{flow_roughening}
\begin{eqnarray}
  \frac d{dl} U_1 &=& (2-\frac {\pi p^2 T}{a^2 \gamma}) U_1,
  \\
  \frac d{dl} \gamma &=& A \frac{2 \pi^4p^2}{\gamma a^2} U_1^2,
\end{eqnarray}
\end{mathletters}
with a temperature dependent coefficient $A$ which is of order one
near the roughening transition.  These equations lead to the phase
diagram sketched in Fig.  \ref{fig_roughening}.

The case with noninteger $p$ involves the analysis of higher order
commensurable states or incommensurable states.  In such cases the
periodic potential is less relevant than in the low order
commensurate cases examined above. Since we find later on that the
periodic potential is irrelevant in the presence of disorder for
integer $p \geq 2$ this will be true also for noninteger $p$. For
more details on commensurate/incommensurate systems the interested
reader is referred to Refs.  \CITE{Bak82,Lyuksyutov}.

\subsection{Pinning potential only ($U=0$)}
\label{sec_V}

In order to discuss the relevance of disorder pinning, we start from
the Hamiltonian ${\cal H}_V$ as given in Eq. (\ref{H_U}), discard
rapidly oscillating terms that are irrelevant on scales much larger
than $a$ and keep only the most relevant term $m=1$ in the sum over
harmonics for the density (see Ref. \onlinecite{Nattermann+00} for
intermediate steps).  After averaging over disorder we find the
effective replica pinning Hamiltonian
\begin{eqnarray}
  {\cal H}_V^{\rm rep} &\simeq& \sum_{\alpha\beta}\int d^2r 
  \left\{-{{\gamma^2\sigma}\over{2 T}}\grad u^\alpha \grad u^\beta
  \right. \nonumber \\ && \left.
    -{{\Delta}\over{a^2 T}}
    \cos {{2\pi}\over a}[u^\alpha(\br)-u^\beta(\br)]\right\}
  \label{H_pin_replica}
\end{eqnarray}
Disorder couples to the $\partial_x u$ term in the density Eq.
(\ref{density}) as a random field and gives rise to the first term in
Eq.  (\ref{H_pin_replica}) with a value $\sigma={{\pi \Delta}\over
  {a^2\gamma^2}}$. (Note that strictly speaking this term should
contain only $\partial_x u^\alpha \partial_x u^\beta$ in the
unrenormalized Hamiltonian; the form written in Eq.
(\ref{H_pin_replica}) anticipates that renormalization generates a
random field coupling to {\em both} components of the gradient.)
Similar to the estimate for the roughening temperature above, one can
estimate the relevance of ${\cal H}_V^{\rm rep}$ from its average with
respect to ${\cal H}_{\rm el}$,
\begin{eqnarray*}
  \langle {\cal H}_V^{\rm rep} \rangle
  &\simeq& -\sum_{\alpha,\beta=1}^NL^2 {{\Delta}\over{a^2 T}} 
  \exp(-2\pi^2\langle[u^\alpha-u^\beta]^2\rangle/a^2)
  \\
  &=&-L^2 N(N-1){\Delta \over{a^2T}} 
  \left({L\over a}\right)^{-{{2\pi}\over a^2}{T\over \gamma}},
\end{eqnarray*}
where we used $\langle u^\alpha(\br) u^\beta(\br) \rangle =
\delta^{\alpha\beta}{1 \over {2\pi}} {T\over \gamma}\ln{L\over a}$.
For temperatures above
\begin{equation}
  T_{\rm SR}={{\gamma a^2}\over{\pi}} 
\end{equation}
the average disorder energy vanishes on large scales. Below, disorder
shows to be relevant and its effects have to be calculated by
renormalization group techniques.  Note that 
\begin{eqnarray}
  T_{\rm SR} \simeq \frac{p^2}{2}T_{\rm R} .
\end{eqnarray}
This relation becomes an identity if the renormalization of $\gamma$
due to the presence of the potentials can be neglected. Then $T_{\rm
  SR}>T_{\rm R}$ for $p>\sqrt 2$.

Cardy and Ostlund\cite{Cardy+82} were the first to derive RG equations
near the transition and Villain and Fernandez\cite{Villain+84} studied
the flow of parameters to their large-scale values at zero
temperature.  A concise summary of these two approaches is given in
Ref. \onlinecite{Nattermann+00}.  We combine the flow equations for
these two temperature ranges by the interpolation
\begin{mathletters}
\label{flow_sr}
\begin{eqnarray}
  \frac{d\sigma}{dl}&=&c_1\frac{a^2\Delta^2}{
    T^2 \gamma^2a^4+\Delta(\gamma^2a^4+c_2\Delta)},
  \\
  \frac{d \Delta}{dl} &=& \left(2-\frac{2\pi T}{\gamma
      a^2}\right)\Delta
  -2\frac{c_2\Delta^2}{\gamma^2a^4+c_2\Delta}.
\end{eqnarray}
\end{mathletters}
The numbers $c_1$ and $c_2$ are of order unity and depend only weakly
on temperature.  $\gamma$ is not renormalized due to a statistical
symmetry,\cite{Hwa+94} just like $\sigma$ does not feed back to
$\Delta$.  This holds for the replica Hamiltonian
(\ref{H_pin_replica}) which is a good approximation on large length
scales.  Smaller scales will weakly renormalize the stiffness $\gamma$
to larger values and generate additional irrelevant terms.  From the
flow equations, $T_{\rm SR}=\gamma a^2/\pi$ is identified as the
temperature above which $\Delta$ is renormalized to zero.
Nevertheless, disorder is marginal for $T>T_{\rm SR}$ since $\sigma$
takes a finite fixed-point value.  Thus, here one has displacement
fluctuations
\begin{equation}
  \overline{\langle u^2 \rangle} = \frac{1}{2\pi}
  \left(\frac{T}{\gamma}+\sigma\right)\ln\frac{L}{a}.
\end{equation}
Below the transition temperature ($T\lesssim T_{\rm SR}$), $\Delta$
flows to a finite fixed-point value and $d\sigma/dl$ becomes constant.
$\sigma$ thus asymptotically has a logarithmic dependence on the scale
$L$ and gives the dominant contribution to the fluctuations
\begin{equation}
  \overline{\langle [u(\br)-u(\bN)]^2 \rangle}
  \sim a^2\chi\ln^2\frac{r}{a}.
  \label{superrough}
\end{equation}
Slightly below $T_{\rm SR}$, $\chi \propto (1-T/T_{\rm SR})^2$.  This
squared-logarithmic roughness (\ref{superrough}) defines the {\em
  superrough} (SR) phase.  In this phase thermal fluctuations can
still give a logarithmic contribution to the correlator
(\ref{superrough}) which -- however -- is sub-dominant.
\begin{figure}
\includegraphics[width=0.9\linewidth]{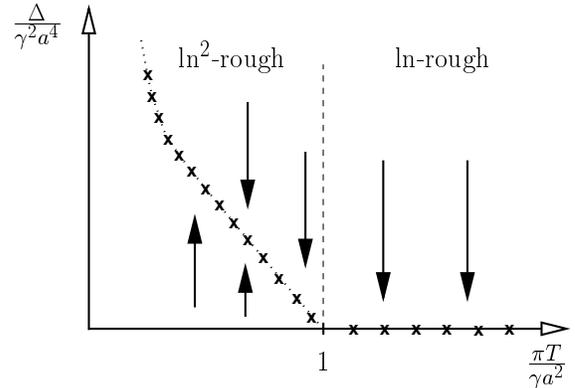}
\caption{Schematic representation of the renormalization group flow
  describing the superroughening transition according to Eqs.
  (\ref{flow_sr}). Arrows indicate the RG flow.  The dashed line is
  the phase boundary between the superrough and rough phase. Crosses
  represent fixed points (note, however, that $\sigma$ flows to
  infinity for $T<T_{\rm SR}=\gamma a^2/\pi$).}
\label{fig_sr}
\end{figure}

\subsection{With complete potential}
\label{sec_combi}

In the presence of both a disorder potential and a periodic pinning
potential it is not immediately clear which one will prevail in what
part of the phase diagram. At first sight it seems to be possible to
have a flat phase, a $\ln$-rough phase, or a $\ln^2$-rough phase.  We
assume that $U$ and $V$ are weak such that a renormalization of the
stiffness $\gamma$ as well as of $T_{\rm R}$ and $T_{\rm SR}$ is
negligible.  Then one always has $T_{\rm R} < T_{\rm SR}$ for $p>\sqrt
2$. Thus, at high temperatures $T>T_{\rm SR}$ the system will be
logarithmically rough since both $U$ and $V$ are irrelevant.  At
intermediate temperatures in the interval $T_{\rm R} <T< T_{\rm SR}$,
$V$ is relevant while $U$ is irrelevant with respect to thermal
fluctuations.  This suggests that the system is superrough.  Although
we ultimately find this to be true (see below), the argument needs to
be refined since it is no longer sufficient that $U$ is irrelevant
with respect to thermal fluctuations.  Instead, we one has to argue
that $U$ is irrelevant at the disorder-dominated, superrough fixed
point.  Eventually, for $T< T_{\rm R}$ one might expect to find a
superrough phase if $U$ is weak compared to $V$ and a flat phase if
$V$ is weak compared to $U$.  However, the following arguments show
that a flat phase is not stable in the presence of disorder and that
superroughness should persist for all $T < T_{\rm SR}$.

In order to argue that a flat phase cannot exist for arbitrarily weak
$V$ and that the stripe array is superrough for all $T< T_{\rm SR}$,
we show (i) that weak periodic potentials ($U \ll V$) are irrelevant
at the superrough fixed point and (ii) that arbitrarily weak disorder
($V \ll U$) will roughen the stripe array even for $T<T_{\rm R}$, i.e.
that disorder is relevant at the flat-phase fixed point, which implies
that $U$ is renormalized to zero and the system is superrough on large
scales.

Consideration (i). For $U \ll V$ the irrelevance of a periodic
potential follows from an analysis in analogy to the one in Sec.
\ref{sec_U}.  We assume that the disorder-induced fluctuations of the
displacement field are Gaussian with a correlator $\overline{\langle
  u^2 \rangle} \sim \frac 12 a^2 \chi \ln^2(L/a) $ [which is implied by
Eq. (\ref{superrough})]. This replaces the thermal correlator in Eq.
(\ref{H_U.av}).  This now leads to
\begin{eqnarray}
  \overline {\langle {\cal H}_U \rangle } \sim 
  - L^2 \frac 1 a \sum_{n \geq 1} U_n
  \left(\frac La \right)^{- \pi^2 n^2 p^2 \chi \ln (L/a)} 
\end{eqnarray}
and the average periodic pinning energy vanishes for large system
sizes.

\begin{figure}
\includegraphics[width=0.9\linewidth]{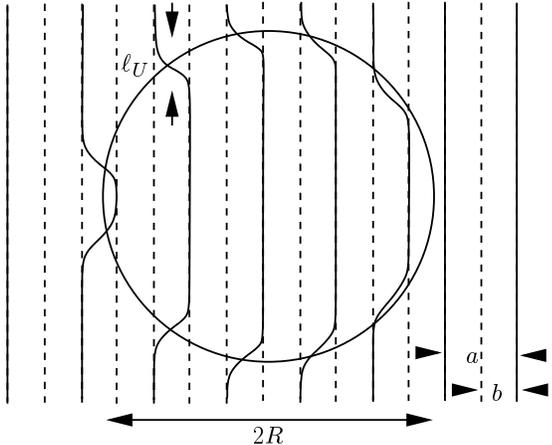}
\caption{Illustration of a soliton of radius $R$ for
  $p=2$.  Full lines represent the strings of holes of average spacing
  $a$, dashed lines are minima of $U$ with spacing $b=a/p$.  The
  soliton width is $\ell_U$.}
\label{fig_soliton}  
\end{figure}

Consideration (ii). The situation $V \ll U$ is more subtle. We assume
that the stripes are locked into a flat state.  We neglect thermal
fluctuations which would renormalize $U$ to smaller values.  For
simplicity of our argument we assume that $p=2$ although the argument
can be generalized to any $p>1$.  In addition, we retain only the most
relevant, lowest harmonic of the periodic pinning energy,
\begin{eqnarray}
  {\cal H}_U \simeq - \frac {U_1}a \int d^2r \ \cos \frac{2 \pi p}a  u(\br).
\end{eqnarray}
In the absence of $V$ all flat states $u(\br)=n b$ with some integer
$n$ would be equivalent. Disorder certainly breaks this degeneracy and
one might expect the stripe array to find a ground state where
$u(\br)$ fluctuates only weakly around $na$ with some particular $n$,
say $n=0$. However, one can show that the ground state is not given by
small fluctuations within this particular ``valley'' $n=0$ but that
solitons (i.e., local areas where $u(\br) \approx \deltan \, b$ with a
shift $\deltan \neq 0$, cf.  Fig.  \ref{fig_soliton}) are preferred
energetically.  The proliferation of a large number of such solitons
implies the irrelevance of $U$ and hence the superroughness of the
stripe array on large scales.

We now examine a disk-like soliton of Radius $R$ and estimate its
elastic energy cost and typical gain of pinning energy in order to
decide whether such solitons are favorable.  From an energetical point
of view the creation of $\deltan=1$ solitons is equivalent to the
creation of a magnetic domain in a random-field Ising model.  For a
strong periodic pinning potential, $U_1 \gg \gamma a$, the soliton has
a narrow border of width $\ell_U \approx \sqrt{a \gamma/U_1} b$ and of
an energy per unit length $\epsilon \approx \sqrt{U_1 a \gamma} /p$.
Thus the elastic energy cost is proportional to the border length,
$E_{\rm el} \propto \epsilon R$.  In the area of the soliton the strings
are exposed to a different disorder potential $V$ (we assume the
disorder correlation length to be small, $\xi \lesssim a$).  Then the
typical energy gain is proportional to the square root of the area
$E_V \propto - \sqrt{ \Delta R^2/a \xi}$.  This gain is larger than the
elastic energy cost for solitons of all sizes for a disorder strength
beyond a threshold value
\begin{equation}
  \Delta_{\rm c} \propto a \xi \epsilon^2 .
\end{equation}
However, for large solitons one has to take into account that the
soliton border will be roughened by disorder. The equivalent
roughening of domain walls in the random field Ising model was studied
in Refs. \CITE{Binder83,Nattermann85}.  As a consequence, the border
line tension $\epsilon$ is renormalized to zero on a finite length
scale
\begin{equation}
  \ell_\epsilon \sim a \exp \left(c \frac{a \xi \epsilon^2}{\Delta}
  \right),
\label{l_eps}
\end{equation}
where $c$ is a constant of order unity.  The creation of solitons of a
size $R$ larger than $\ell_\epsilon$ is thus energetically favorable.
Overlapping solitons of unbounded size imply a roughness of the stripe
array, provided the sum of the shifts $\sum_i \deltan_i$ at a given
position increases with the number of solitons (enumerated by the
index $i$) that include this site.  In principle, the interaction
between solitons could lead to a compensation of the shifts between
pairs of solitons.  However, the interaction between the solitons is
short-ranged on the scale $\ell_U$ and it cannot compete with a
disorder energy that discriminates between a shift $\deltan_i=1$ and a
$\deltan_i=-1$ for each soliton.  Although the pinning energy of these
two states is identical in the bulk area, the disorder energy of these
two solitons is different in the border region and leads to an energy
contribution proportional to $\sqrt R$.  Therefore the multiple
creation of large and overlapping solitons leads to uncorrelated
contributions to shifts $\deltan$ and therefore implies the roughness
of the stripe array on scales beyond $\ell_\epsilon$.  In this sense
$U$ is irrelevant on large scales in the presence of arbitrarily weak
$V$, although on small scales the stripes will be confined to valleys
of $U$.  Since $U$ is irrelevant, the stripe array will be superrough
as in the absence of $U$.

Although the case $p=1$ appears not to be of physical relevance for
striped systems, we add as a side remark that for $p=1$ a flat phase
can exist. In this case the elastic energy cost $\propto R$ of a
soliton cannot be compensated by a disorder energy which no longer has
a bulk contribution $\propto R$ but only a border contribution
$\propto \sqrt R$.

Strictly speaking, we have shown the irrelevance of $U$ only for
integer values of $p$. Since noninteger rational values of $p$ would
correspond to commensurabilities of higher order, they are even more
susceptible to the destruction of long-range order by $U$.

It is interesting to note that the absence of a flat phase for $p>1$
is peculiar to two dimensions. In dimensions $2<d<4$ a flat phase is
stable for disorder weaker than a threshold
value.\cite{Emig+98,Emig+99} Therefore, a stack of planar stripe
arrays with a finite coupling between the planes will exhibit also a
flat phase for a disorder strength below a certain threshold value.

\section{Adding quantum fluctuations}
\label{sec_quant}

Now we finally study how the potentials $U$ and $V$ affect the elastic
string model taking into account its quantum mechanical nature.  In
order to analyze whether these potentials are relevant at all, the
scaling arguments used in section \ref{sec_class} can be applied
analogously to the dynamic action.  Since the displacement
correlations of the unpinned array are qualitatively different for
$T=0$ (where the array is flat) and $T>0$ (where the array is rough)
these cases will be discussed separately.

\subsection{At zero temperature}

In this section we focus on the case with disorder but without a
periodic potential.  Before we turn to the analysis of this case, it
is worthwhile to point out that this system is essentially a
two-dimensional ``Bose glass.''  Because of an analogy between a
$d$-dimensional bosonic problem at zero temperature and a classical
$(d+1)$-dimensional problem a finite temperature,\cite{FisherMPA+89}
superconducting vortices in the presence of columnar pinning centers
provide another equivalent system that has received a lot of recent
attention.\cite{Nelson+92,Nelson+93} The problem at hand, a bosonic
one-component displacement field at $T=0$, is equivalent to a stack of
classical elastic layers which was studied by Balents.\cite{Balents93}
According to his analysis -- which is restricted to dimensions close
to $d=4$ -- the displacement field is logarithmically rough.  We now
focus on two dimensions because characteristic modifications leading to
superroughness are to be expected.

The analysis of the disorder pinning is most convenient using the
replicated action
\begin{eqnarray}
  {\cal S}_V^{\rm rep}&\simeq& \sum_{\alpha\beta} \int d\tau 
  d\tau'  d^2r 
  \left\{-{{\gamma^2\sigma}\over{2 \hbar^2}} \grad u^\alpha(\br,\tau) 
    \grad u^\beta(\br,\tau')
  \right. \nonumber \\ && \left.
    -{{\Delta}\over{a^2 \hbar^2}} 
    \tDelta [u^\alpha(\br,\tau)-u^\beta(\br,\tau')]\right\}
\label{S_V.rep}
\end{eqnarray}
where 
\begin{eqnarray}
  \tDelta(u)=\cos {{2\pi}\over a} u
\end{eqnarray}
if we retain only the lowest harmonic of the stripe density as most
relevant term as in Eq. (\ref{H_pin_replica}). 

For a scaling analysis of this action contribution, we consider a
rescaling of space, time, displacement and action quantum according to
\begin{eqnarray}
  \br =  e^l  \br_l, \quad
  \tau = e^{zl}  \tau_l , \quad
  u= e^{\zeta l} u_l, \quad
  \hbar = e^{\eta l} \hbar_l 
\end{eqnarray}
with a dynamical exponent $z$, roughness exponent $\zeta$ and action
scaling exponent $\eta$. In order to analyze the relevance of quantum
fluctuation later on we allow for a rescaling of $\hbar$.  We note that
due to a statistical tilt symmetry the flow equation of $\gamma$
consists only of the scaling part \cite{Hwa+94}
\begin{eqnarray}
  \frac d{dl} \gamma = (z-\eta + 2 \zeta) \gamma 
\end{eqnarray}
which implies
\begin{eqnarray}
  \eta=z+2\zeta
\label{eta_rel}
\end{eqnarray}
at any possible fixed point.  If $\eta>0$ quantum fluctuations are
irrelevant on large scales and the fixed point can be called a
``classical'' ($\hbar_\infty=0$) fixed point in analogy to the
irrelevance of thermal fluctuations for classical
systems.\cite{FisherDS86}

To establish the relevance of pinning we note that the action of the
unpinned system ${\cal S}_{\rm el}$ is invariant under a rescaling
with $z=1$, $\eta=0$ and $\zeta=-\frac 12$.  According to Eq.
(\ref{corr_u_el}) the displacement fluctuations are finite for $T=0$
(this is reflected by the negative value of $\zeta$) which implies
that $\langle \cos {{2\pi} \over a}[u^\alpha(\br,\tau) -
u^\beta(\br,\tau')] \rangle$ is finite since in the unpinned system
$\langle [u^\alpha(\br,\tau) - u^\beta(\br,\tau')]^2 \rangle \simeq
\hbar/({a \sqrt{\pi \gamma \mu}})$ for $l\to \infty$ and thus disorder
is strongly relevant, $\langle {\cal S}_V^{\rm rep} \rangle \propto
e^{4l}$.

Disorder can be taken into account on a crude level (the ``random
field'' approximation) by retaining in action (\ref{S_V.rep}) only the
harmonic parts bilinear in $u$.  Then this action part is found to be
scale invariant together with ${\cal S}_{\rm el}$ for $z=1$,
$\zeta=1$, and $\eta=3$.  To gain qualitative insight into the
spatio-temporal correlations we calculate the displacement
fluctuations in this random-field approximation.  We find
\begin{mathletters}
\begin{eqnarray}
  \overline{\langle
  [u^\alpha(\br,\tau)-u^\beta(\bN,0)]^2 \rangle} =  C_1^{\alpha
  \beta}(\br,\tau) + C_2(\br)
\label{corr.u.rf}
\end{eqnarray}
with a contribution $C_1$ from quantum fluctuations and a disorder
contribution $C_2$.  $C_1^{\alpha \beta}(\br,\tau)$ vanishes for
$\alpha = \beta$ and $\br=\bN$ and $\tau=0$ and takes a finite value
\begin{eqnarray}
  C_1^{\alpha \beta}(\br,\tau) \simeq \frac{\hbar}{a\sqrt{\pi \gamma \mu}}
\end{eqnarray}
for $\alpha \neq \beta$ or $r \gg a$ or $\tau \gg \sqrt{\mu/\gamma}
a$.  $C_2(\br)$ is rough; its roughness
\begin{eqnarray}
  C_2(\br) \simeq \frac \sigma \pi \ln \frac ra + \frac{4 \pi
    \Delta}{\gamma^2 a^4} r^2
  \label{C_2}
\end{eqnarray}
\end{mathletters}
is dramatically overestimated in this approximation. From this
correlation function we recognize that while disorder roughens the
displacement in {\em spatial} directions (consider large $r$ for
$\tau=0$), it preserves the flatness in {\em temporal} directions
(consider large $\tau$ for $r=0$).  These qualitative properties
should hold even after renormalization effects due to the anharmonic
terms in action (\ref{S_V.rep}) are taken into account.

Although a systematic renormalization group analysis is very intricate
and beyond the scope of this article, we present arguments in favor of
$\eta>0$ at the true fixed point.  First, we use the fact that a fixed
point with a periodic correlator $\tDelta(u)$ can exist only for
$\zeta=0$.  Thus
\begin{eqnarray}
  \eta=z
\end{eqnarray}
from Eq. (\ref{eta_rel}) and it is sufficient to show $z>0$ for the
irrelevance of quantum fluctuations. It is natural to assume that the
dynamic exponent is positive, if not diverging on large scales as is
typical of glassy systems where the dynamics is governed by tunneling
through divergent barriers.

The difficulty\cite{Balents93} to calculate $z$ is related to the fact
that $\tDelta(u)$ may become nonanalytic near $u=0$.  While such a
nonanalyticity follows from a functional renormalization group
analysis near $d=4$,\cite{FisherDS86} it is not clear whether such a
nonanalyticity is present in $d=2$.  According to the analysis
summarized in Sec.  \ref{sec_V}, such a nonanalyticity is absent at
finite temperature near $T_{\rm SR}$.  At zero temperature, the strong
coupling analysis of Villain and Fernandez\cite{Villain+84} misses
such a possible nonanalyticity.  Therefore we consider both
possibilities.

We first assume analyticity of the function $\tDelta(u)$. In this case
the ``dynamic stiffness'' will be renormalized according to
\begin{eqnarray}
  \frac d{dl} \mu = (2-z+2\zeta-\eta) \mu + \beta_\mu[\tDelta].
  \label{flow.mu}
\end{eqnarray}
The functional $\beta_\mu[\tDelta]$ represents the vertex corrections
arising from the ${\cal S}_V^{\rm rep}$. These corrections are
expected to be positive since disorder pins the stripes at minima of
$V$, thereby confining the temporal fluctuations which amounts to an
increase of the renormalized $\mu$. If $\beta_\mu[\tDelta]$ is finite
at the fixed point, then $\eta=z=1+\beta_\mu[\tDelta]/(2\mu^*)$
($\mu^*$ denotes the fixed-point value of $\mu$).  Then $\eta>0$ and
the fixed point will be classical.

In case the fixed-point correlator is nonanalytic the functional
$\beta_\mu[\tDelta]$ diverges.\cite{Balents93} This signals a
qualitative increase of the dynamic stiffness, which should be
described by a kinetic term of the form\cite{Balents93}
\begin{eqnarray}
  {\cal S}_{\rm kin} = \frac 1 \hbar 
  \int_0^{\hbar/T} d\tau  \int d^2r \ \nu |\partial_\tau u|
  \label{S_kin_nu}
\end{eqnarray}
which is more relevant than the original kinetic term in Eq.
(\ref{def.S}). The coefficient $\nu$ would flow according to
\begin{eqnarray}
   \frac d{dl} \nu = (2+\zeta-\eta) \nu + \beta_\nu[\tDelta]
  \label{flow.nu}
\end{eqnarray}
with another functional $\beta_\nu[\tDelta]>0$. This would imply an
even larger dynamical exponent $z=\eta=2+\beta_\nu[\tDelta]/\nu^*$ and
even stronger irrelevancy of quantum fluctuations.

Thus, in any case $\eta>0$ and the system flows to the classical fixed
point value for $V \neq 0=U$. Quantum effects will result only in a
finite renormalization of the parameters in the classical system.  The
most important renormalization effect concerns an increase of the
dynamic stiffness with a possible generation of $\nu$.  Although there
is no way of handling a kinetic action of the form (\ref{S_kin_nu}),
we expect quantum fluctuations on small scales to induce a flat but
finite quantum contribution $C_1$ to the displacement correlation.
The classical contribution $C_2$ will be renormalized as in the
absence of quantum fluctuations [the proper correlation can be
obtained from equation (\ref{C_2}) by inserting the scale-dependent
values of $\Delta$ and $\sigma$ as obtained from the flow equations
(\ref{flow_sr}) without rescaling; in the superrough phase $\sigma
\propto \ln r$ and $\Delta \propto (\ln r)/r^2$].

Since the presence of the disorder potential implies the irrelevance
of quantum fluctuations, they cannot be expected to modify the
competition between $U$ and $V$ as was analyzed in Sec.
\ref{sec_combi}.  Thus, the quantum array has superrough spatial
correlations at $T=0$.

\subsection{At finite temperature}

An inspection of the correlator (\ref{corr_u_el}) of the unpinned
system suggests that thermal fluctuations dominate over quantum
fluctuations on large scales.  In fact, quantum fluctuations are
irrelevant at the classical fixed point also for $T>0$.

This can be seen from the action as follows. The classical fixed
points (with both thermal roughness or superroughness) are described
by $\zeta=0$. The finiteness of the time integral implies $z=0$ and
according to Eq.  (\ref{eta_rel}) also $\eta=0$.  Then the effective
``dynamical stiffness'' flows to infinity according to Eq.
(\ref{flow.mu}) or Eq. (\ref{flow.nu}). Thereby temporal fluctuations
of the displacement are suppressed on large scales, on which the
system is described by the static limit
\begin{eqnarray}
  {\cal S} \to \frac 1T {\cal H}.
\end{eqnarray}
Thus quantum fluctuations will lead only to a renormalization of the
parameters in the classical description.  Therefore the scaling
arguments that were applied in section \ref{sec_class} to the
Hamiltonian hold also for the dynamic action.  Thus the system in
disorder will be superrough at low temperatures -- without or with an
additional periodic potential -- while it will be thermally rough at
high temperatures.

\section{Summary and Discussion}
\label{sec_disc}

So far, we have analyzed the stripe array in the elastic
approximation, i.e. neglecting dislocations. Before we discuss the
relevance of dislocations, we summarize our results.  In general, we
found quantum fluctuations to be irrelevant in the presence of thermal
and/or disorder-induced fluctuations, i.e., to renormalize the
classical elastic model only weakly.

In the absence of disorder ($\Delta=0$), the stripe array locks into a
commensurate periodic potential below a roughening temperature $T_{\rm
  R}=2 \gamma b^2/\pi = 2 \gamma a^2 /\pi p^2$.  This transition
temperature is proportional to the stiffness $\gamma$ which is
implicitly temperature dependent due to an entropic contribution
$\gamma_{\rm entr} = \pi T/a^2$.\cite{Pokrovsky+79} If there were only
this entropic contribution, the system would always be unlocked
($T>T_{\rm R}$) at finite temperatures for $p>\sqrt2$. However, a
temperature independent contribution to $\gamma$, which arises from
the Coulomb interaction, leads to a lock-in transition at a finite
temperature for $p>\sqrt2$, where the translational order changes from
long ranged to quasi-long ranged (with logarithmic roughness).

The large-scale structure of the stripe array is dramatically
influenced by the presence of disorder ($\Delta>0$). If there were no
periodic potential, the stripe array would undergo a superroughening
transition at $T_{\rm SR}= \gamma a^2/\pi$. For $T>T_{\rm SR}$ the
array would be unpinned with $\ln$-roughness, while it would be pinned
and superrough for $T<T_{\rm SR}$.  The same scenario holds also in
the presence of the periodic potential, from which the stripe array
always unlocks (on sufficiently large scales even for arbitrarily weak
disorder).  Therefore, the array is {\em rough at all temperatures}
for $\Delta>0$.  However, for weak disorder and strong periodic
potential the crossover length scale from flat to superrough
correlations will be exponentially large, cf.  Eq. (\ref{l_eps}).

Hasselmann {\em et al.}\cite{Hasselmann+99} previously proposed a
phase diagram for a single stripe with flat and disordered phases. To
our understanding, the disorder was effectively assumed to have
long-ranged correlations, which allows for the existence of a flat
phase even for a single stripe.  In contrast to this we consider
disorder with short-ranged correlations and find that it always
roughens the stripe array.  Because a single stripe represents an
elastic system of lower dimensionality than the stripe array, our
finding implies also the roughness of a single stripe in disorder with
short-ranged correlations.

Now we come back to discuss the relevance of topological defects in
the stripe array, starting with the simplest situation for $U=V=0$.
At low temperatures the stripe array can be considered as a
``smectic'' with quasi-long-range translational order and long-range
orientational order. At a temperature
\begin{eqnarray}
  T_{\rm m}= \frac{\gamma a^2}{8\pi}
\end{eqnarray}
it would melt\cite{Kosterlitz+73} due to a proliferation of
dislocations into a ``nematic'' liquid with short-range translational
order and quasi-long-range orientational order, before the
proliferation of disclinations drives a second transition into an
isotropic liquid with short-ranged orientational order.  While this
scenario is well known for classical systems (for a review see e.g.
Ref. \CITE{Nelson83}), its relevance to doped Mott insulators was
pointed out by Kivelson, Fradkin, and Emery.\cite{Kivelson+98}

It is instructive to compare the melting temperature to the other
characteristic temperatures related to the potentials.  As pointed out
by Jos\'e {\em et al.},\cite{Jose+77} distinct melting and lock-in
transitions can exist only for $p \geq 4$ since $T_{\rm R}=(4/p)^2
T_{\rm m}$.  For $p<4$ they will merge to a single phase transition.
The effect of disorder is quite virulent: In the elastic
approximation, it makes the system superrough at temperatures below
$T_{\rm SR} = 8 T_{\rm m}$.  However, superroughness implies that
dislocations become energetically favorable [this follows from the
flow equations (\ref{flow_sr}), see. e.g. Ref. \CITE{Ledoussal+00}].
Therefore, their density will be finite even at zero temperature.
Nevertheless, for weak disorder and at low temperatures the length
scale where free dislocation appear can be extremely large because of
a $U$ that tends to lock the stripes up to exponentially large scales.
Since melting in the absence of disorder occurs at a temperature,
$T_{\rm m}<T_{\rm SR}$, free dislocations will be present at {\em all}
temperatures.

The presence of dislocations reduces the quasi-long-range order (where
the satellite peaks have an algebraic singularity due to the
$\ln$-roughness; $\ln^2$-roughness corresponds to an algebraic
singularity with an exponent that depends weakly on the wave vector)
to short-range order with a correlation length of the order of the
distance between (free) dislocations.  Thus, the saturation of the
dislocation density at low temperatures implies also the saturation of
the correlation length at a maximum value.  This conclusion is
consistent with experimental observations in the
cuprates\cite{Tranquada+99} (here it was observed in magnetic
ordering, which can be destroyed due to a coupling to disorder even
without a distortion of the stripe array) as well as in the
nickelates\cite{Lee+97,Du+00} (here the correlation length is
explicitly that of charge order).

Note that the {\em true} correlation length of charge order $\xi^{\rm
  C}$ (which we identify with the distance between free dislocations)
can be related to the measured width of peaks in the structure
function only if the wave-vector resolution is much smaller than
$1/\xi^{\rm C}$.  Thus, a system without dislocations (with $\ln$- or
$\ln^2$-roughness) has an infinite correlation length and the {\em
  apparent} correlation length deduced from experiments would be
resolution limited.  Zachar\cite{Zachar00} recently proposed an
explanation of the observed apparent correlation lengths in terms of
chaotic fluctuations of the distance between neighboring stripes (see
also Ref.  \CITE{Tranquada+99}), excluding explicitly a key role of
dislocations.  However, this argument is based on the assumption of
noninteger $p$ and cannot account for the finite correlation length
observed for integer $p$ in the
nickelates\cite{Lee+97,Du+00,note_zachar} and the
cuprates.\cite{Yamada+98} It also cannot explain why the transverse
correlation length (along the stripes) -- which apparently is not
resolution limited\cite{Niemoeller+99} -- can be finite.  Even more,
the fact that longitudinal and transverse correlation lengths are
roughly of the same size\cite{Zimmermann+98,Niemoeller+99} is
consistent with their relation to the dislocation density.

As both the roughening transition is washed out by disorder and the
superroughening transition is washed out by the presence of
dislocations, no sharp transition will exist in the thermodynamic
limit.  Nevertheless, crossover phenomena may be observable.  For weak
disorder and/or strong lattice potential the stripe array appears to
be locked to the lattice potential up to very large length scales.  It
``unlocks'' also on finite scales with increasing temperature or
decreasing lattice potential.  An apparent lock-in transition was
observed\cite{Tranquada+95} in La$_{2-x-y}$Nd$_y$Sr$_x$CuO$_4$ where
the strength of the lattice potential corresponds to the Nd
concentration.  Since $T_{\rm SR} \gg T_{\rm m}$ the density of free
dislocations will be high near $T_{\rm SR}$ so that reminiscences on
finite scales of the superroughening transition are unlikely to
survive.

Since the array has short-range order at all temperatures for
$\Delta>0$ we expect the absence of commensurate/incommensurate
transitions (identified from fluctuations on asymptotically large
length scales).  Nevertheless, the experimental observations of
anomalies at particular values of $p$ can be related to the effects of
$U$ on finite length scales.

In view of our conclusion about the absence of commensuration effects
we comment on the $\delta=\frac 18$ problem, i.e., the observation
that $\delta=1/2p$ apparently saturates at $\delta=\frac 18$ near a
doping $x=0.125$ of the cuprates.  If this were a commensuration
effect as suggested previously, \cite{Tranquada+95,Nayak+96} one would
expect to observe the value $\delta=\frac 18$ around $x=\frac 18$,
i.e.  above ($x>\delta$) and below ($x<\delta$) the matching
$x=\delta$.  To the best of our knowledge (cf. the data collected in
Fig. 7 of Ref.  \CITE{Yamada+98}), there is no evidence for data with
$x<\delta$.  As pointed out by Yamada {\em et al.},\cite{Yamada+98}
the saturation of $\delta$ coincides with a saturation of the
effective hole concentration in the CuO$_2$ planes beyond a certain
doping level $x \approx 0.12$.  Thus it is conceivable that other
mechanisms limit the effective hole concentration and lead to a
plateau in $\delta$.

In conclusion, we have pointed out the relevance of disorder for a
stripe array even in states where its period is commensurate with the
atomic structure. We found that on large scales pinning by disorder
dominates over pinning by the atomic structure.  This induces the
superroughness of the array and, on sufficiently large scales, the
presence of free dislocations even at low temperatures, which explains
the saturation of the correlation length observed in experiments.

\section*{Acknowledgments}

The authors are grateful to T.  Nattermann for stimulating discussions
and valuable hints.  They acknowledge financial support by Deutsche
Forschungsgemeinschaft through SBF341.


\begin{thebibliography}{10}

\bibitem{Schulz89}
H.~J. Schulz, J. Physique {\bf 50},  2833  (1989).

\bibitem{Zaanen+89}
J. Zaanen and O. Gunnarsson, Phys. Rev. B {\bf 40},  7391  (1989).

\bibitem{Emery+90}
V.~J. Emery, S.~A. Kivelson, and H.-Q. Lin, Phys. Rev. Lett. {\bf 64},  475
  (1990).

\bibitem{Cheong+91}
S.-W. Cheong {\it et~al.}, Phys. Rev. Lett. {\bf 67},  1791  (1991).

\bibitem{Mason+92}
T.~E. Mason, G. Aeppli, and H.~A. Mook, Phys. Rev. Lett. {\bf 68},  1414
  (1992).

\bibitem{Tranquada+95}
J.~M. Tranquada {\it et~al.}, Nature (London) {\bf 375},  561  (1995).

\bibitem{Dai+98}
P. Dai, H.~A. Mook, and F. Dogan, Phys. Rev. Lett. {\bf 80},  1738  (1998).

\bibitem{Hayden+92}
S.~M. Hayden {\it et~al.}, Phys. Rev. Lett. {\bf 68},  1061  (1992).

\bibitem{Chen+93}
C.~H. Chen, S.-W. Cheong, and A.~S. Cooper, Phys. Rev. Lett. {\bf 71},  2461
  (1993).

\bibitem{Tranquada+94}
J.~M. Tranquada, D. Buttrey, V. Sachan, and J.~E. Lorenzo, Phys. Rev. Lett.
  {\bf 73},  1003  (1994).

\bibitem{Lee+97}
S.-H. Lee and S.-W. Cheong, Phys. Rev. Lett. {\bf 79},  2514  (1997).

\bibitem{Mook+00}
H.~A. Mook, P. Dai, F. D\u{o}gan, and R.~S. Hunt, Nature (London) {\bf 404},
  729  (2000).

\bibitem{Kivelson+96}
S.~A. Kivelson and V.~J. Emery, Synth. Met. {\bf 80},  151  (1996).

\bibitem{Zachar+98}
O. Zachar, S.~A. Kivelson, and V.~J. Emery, Phys. Rev. B {\bf 57},  1422
  (1998).

\bibitem{Emery+93}
V.~J. Emery and S.~A. Kivelson, Physica C {\bf 289},  597  (1993).

\bibitem{Mook+99}
H.~A. Mook and F. D\u{o}gan, Nature (London) {\bf 401},  145  (1999).

\bibitem{Tranquada+97}
J.~M. Tranquada {\it et~al.}, Phys. Rev. Lett. {\bf 78},  338  (1997).

\bibitem{Kimura+99}
H. Kimura {\it et~al.}, Phys. Rev. B {\bf 59},  6517  (1999).

\bibitem{Crawford+91}
M.~K. Crawford {\it et~al.}, Phys. Rev. B {\bf 44},  7749  (1991).

\bibitem{Nakamura+92}
Y. Nakamura and S. Uchida, Phys. Rev. Lett. {\bf 46},  5841  (1992).

\bibitem{Emery+97}
V.~J. Emery, S.~A. Kivelson, and O. Zachar, Phys. Rev. B {\bf 56},  6120
  (1997).

\bibitem{Zaanen+96}
J. Zaanen, M.~L. Horbach, and W. van Saarloos, Phys. Rev. B {\bf 53},  8671
  (1996).

\bibitem{MSmith+98}
C. {Morais Smith}, Y. Dimashko, N. Hasselmann, and A.~O. Caldeira, Phys. Rev. B
  {\bf 58},  453  (1998).

\bibitem{Nayak+96}
C. Nayak and F. Wilczek, Int. J. Mod. Phys. B {\bf 10},  2125  (1996).

\bibitem{Yamada+98}
K. Yamada {\it et~al.}, Phys. Rev. B {\bf 57},  6165  (1998).

\bibitem{Moodenbaugh+88}
A.~R. Moodenbaugh, Y.~X.~M. Suenaga, T.~J. Folkerts, and R.~N. Shelton, Phys.
  Rev. B {\bf 38},  4596  (1988).

\bibitem{note_comm}
Note that our usage of the notions commensurate and incommensurate follows the
  traditional usage in the context of commensurate-incommensurate
  transitions.\cite{Bak82,Lyuksyutov} We distinguish {\em incommensurability}
  from a {\em mismatch} $\delta$ of Bragg peaks in the structure factor, whreas
  in the stripe literature incommensurability is used synonymously to mismatch.
  Our notions allow to distinguish incommensurate and commensurate states with
  finite mismatch.

\bibitem{Cheong+94}
S.-W. Cheong {\it et~al.}, Phys. Rev. B {\bf 49},  7088  (1994).

\bibitem{Ramirez+96}
A.~P. Ramirez {\it et~al.}, Phys. Rev. Lett. {\bf 76},  447  (1996).

\bibitem{Tranquada+99}
J.~M. Tranquada, N. Ichikawa, and S. Uchida, Phys. Rev. B {\bf 59},  14712
  (1999).

\bibitem{Du+00}
C.-H. Du {\it et~al.}, Phys. Rev. Lett. {\bf 84},  3911  (2000).

\bibitem{Hasselmann+99}
N. Hasselmann, A.~H.~C. Neto, C. {Morais Smith}, and Y. Dimashko, Phys. Rev.
  Lett. {\bf 82},  2135  (1999).

\bibitem{Zaanen00}
J. Zaanen, Phys. Rev. Lett. {\bf 84},  753  (2000).

\bibitem{Nattermann+00}
T. Nattermann and S. Scheidl, Adv. Phys. {\bf 49},  607  (2000).

\bibitem{Nozieres92}
P. Nozi\`eres,  in {\em Solids far from Equilibrium}, edited by C. Godr\`eche
  (Cambridge Univ. Press, Cambridge, 1992).

\bibitem{Bak82}
P. Bak, Rep. Prog. Phys. {\bf 45},  587  (1982).

\bibitem{Lyuksyutov}
I. Lyuksyutov, A.~G. Naumovets, and V. Pokrovsky, {\em Two-dimensional
  crystals} (Academis Press, San Diego, 1992).

\bibitem{Cardy+82}
J.~L. Cardy and S. Ostlund, Phys. Rev. B {\bf 25},  6899  (1982).

\bibitem{Villain+84}
J. Villain and J.~F. Fernandez, Z. Phys. B {\bf 54},  139  (1984).

\bibitem{Hwa+94}
T. Hwa and D. Fisher, Phys. Rev. Lett. {\bf 72},  2466  (1994).

\bibitem{Binder83}
K. Binder, Z. Phys. B {\bf 50},  343  (1983).

\bibitem{Nattermann85}
T. Nattermann, Phys. Stat. Sol. (b) {\bf 132},  125  (1985).

\bibitem{Emig+98}
T. Emig and T. Nattermann, Phys. Rev. Lett. {\bf 81},  1469  (1998).

\bibitem{Emig+99}
T. Emig and T. Nattermann, Eur. Phys. J. B {\bf 8},  525  (1999).

\bibitem{FisherMPA+89}
M.~P.~A. Fisher and D.~H. Lee, Phys. Rev. B {\bf 39},  2756  (1989).

\bibitem{Nelson+92}
D.~R. Nelson and V.~M. Vinokur, Phys. Rev. Lett. {\bf 68},  2398  (1992).

\bibitem{Nelson+93}
D.~R. Nelson and V.~M. Vinokur, Phys. Rev. B {\bf 48},  13060  (1993).

\bibitem{Balents93}
L. Balents, Europhys. Lett. {\bf 24},  489  (1993).

\bibitem{FisherDS86}
D.~S. Fisher, Phys. Rev. Lett. {\bf 56},  1964  (1986).

\bibitem{Pokrovsky+79}
V.~L. Pokrovsky and A.~L. Talapov, Phys. Rev. Lett. {\bf 42},  65  (1979).

\bibitem{Kosterlitz+73}
J.~M. Kosterlitz and D.~J. Thouless, J. Phys. C {\bf 6},  1181  (1973).

\bibitem{Nelson83}
D.~R. Nelson,  in {\em Phase Transitions and Critical Phenomena}, edited by C.
  Domb and J.~L. Lebowitz (Academic Press, London, 1983), Vol.~7.

\bibitem{Kivelson+98}
S.~A. Kivelson, E. Fradkin, and V.~J. Emery, Nature (London) {\bf 393},  550
  (1998).

\bibitem{Jose+77}
J.~V. Jos\'{e}, L.~P. Kadanoff, S. Kirkpatrick, and D.~R. Nelson, Phys. Rev. B
  {\bf 16},  1217  (1977).

\bibitem{Ledoussal+00}
P. {Le Doussal} and T. Giamarchi, Physica C {\bf 331},  233  (2000).

\bibitem{Zachar00}
O. Zachar, Phys. Rev. B {\bf 62},  13836  (2000).

\bibitem{note_zachar}
Note also that in the nickelates the charge correlation length is {\em larger}
  than the spin correlation length by roughly a factor 4,\cite{Lee+97}, while
  it should be {\em smaller} in the absence of topological defects according to
  Ref. \CITE{Zachar00}.

\bibitem{Niemoeller+99}
T. Niem\"oller {\it et~al.}, Eur. Phys. J. B {\bf 12},  509  (1999).

\bibitem{Zimmermann+98}
{Von Zimmermann} {\it et~al.}, Europhys. Lett. {\bf 41},  629  (1998).

\end{thebibliography}

\end{multicols}

\end{document}